\documentclass{svjour3}                     
\smartqed  
\usepackage{amssymb,amsfonts}
\usepackage{latexsym}
\usepackage{amscd}
\usepackage{amsmath}
\usepackage{graphicx}
\journalname{ECCS 2009}

\newcommand{\gh}{1ex}
\newcommand{\one}{\includegraphics[height=\gh]{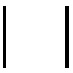}}
\newcommand{\two}{\includegraphics[height=\gh]{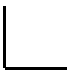}}
\newcommand{\three}{\includegraphics[height=\gh]{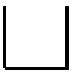}}
\newcommand{\four}{\includegraphics[height=\gh]{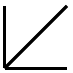}}

\newcommand{\six}{\includegraphics[height=\gh]{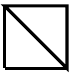}}
\newcommand{\seven}{\includegraphics[height=\gh]{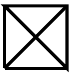}}

\newcommand{\nine}{\includegraphics[height=\gh]{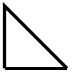}}

\newcommand{\be}{\begin{equation}}
\newcommand{\ee}{\end{equation}}
\newcommand{\bes}{\begin{eqnarray}}
\newcommand{\ees}{\end{eqnarray}}
\newcommand{\besn}{\begin{eqnarray*}}
\newcommand{\eesn}{\end{eqnarray*}}

\newcommand{\abs}[1]{\left\lvert#1\right\rvert} 

\newcommand{\Ha}{\mathcal{H}} 
\newcommand{\comment}[1]{ }

\newcommand{\ie}{i.e.\;}  

\newcommand{\ER}{Erd\H{o}s-R\'{e}nyi }
\renewcommand{\subset}{\subseteq}  

\newcommand{\defas}{\mathrel{\mathop{:}}=}   
\newcommand{\set}[1]{\left\lbrace #1 \right\rbrace} 

\begin{document}

\title{Quantifying structure in networks \thanks{Supported by the Volkswagen Foundation}
}


\author{Eckehard Olbrich \and Thomas Kahle \and Nils Bertschinger \and Nihat Ay \and J\"urgen Jost}

\authorrunning{Olbrich et al.} 

\institute{Eckehard Olbrich \and Thomas Kahle \and Nils Bertschinger \and Nihat Ay \and J\"urgen Jost 
\at Max Planck Institute for Mathematics in the Sciences, Leipzig, Germany
\and 
Nihat Ay \and J\"urgen Jost 
\at Santa Fe Institute, 1399 Hyde Park Road, Santa Fe, New Mexico 87501, USA}

\date{Received: date / Accepted: date}

\maketitle

\begin{abstract}
  We investigate exponential families of random graph distributions as
  a framework for systematic quantification of structure in
  networks. In this paper we restrict ourselves to undirected
  unlabeled graphs. For these graphs, the counts of subgraphs with no
  more than $k$ links are a sufficient statistics for the exponential
  families of graphs with interactions between at most $k$ links. In
  this framework we investigate the dependencies between several
  observables commonly used to quantify structure in networks, such as
  the degree distribution, cluster and assortativity coefficients.
\end{abstract}

\section{Introduction}
\label{intro}
The notion of ``complex networks''  is usually utilized 
in an informal manner, intending to suggest that these networks are not simple in
some sense or another. Among the simple networks
one would include regular lattices on the
one side and purely random networks, i.e. \ER random graphs
or Bernoulli graphs, on the other side. In contrast, in the physics
literature, two types of networks are considered as prototypes of complex
networks: scale free graphs, i.e. graphs with degrees distributed
according to a power law and small world networks, i.e. graphs with a
small diameter, but higher cluster coefficient than a 
Bernoulli graph with the same diameter. Thus the degree distribution
and the cluster coefficient are used to define certain kinds of
complex networks. Another structural property, utilized to assess whether some graph should
be termed complex, is its assortativity or disassortativity, describing whether high
degree nodes are more often connected to high degree nodes or to low
degree nodes. 

These properties are not independent of each other;
the degree distribution, for instance, imposes constraints on the
degree of assortativity. Therefore it would be desirable to have a
general framework that allows to study these dependencies in a
systematic way and, in particular, to quantify structure
and therefore complexity of networks. Here we propose
the theory of hierarchically structured exponential families
\cite{Amari2001,Kahle2009} with the help of which we can, starting from the \ER random
graphs, incorporate more and more interaction between parts of a
network and provide a framework for quantifying the degree of
interaction.

In this theory ``complexity'' means statistical complexity
\cite{Ay2006b}: In order to distinguish between the structure and the
random part, not only one object is considered, but a set of objects
equipped with a probability distribution. ``Random'' then means
statistically independent. Accordingly, measures of statistical
complexity quantify statistical dependencies in a
distribution. They vanish in both cases of a totally ordered and a
totally random system.

Using the notion of ``statistical complexity'' to characterize single
objects such as a given network is problematic in the
following sense. If we speak about the complexity of a single
network, we have to consider it as typical in an ensemble
of networks. This assumption need not always be justified. If it is
satisfied, however, we can use an
ergodicity-type argument to approximate ensemble means by counts over
a single typical instance in the ensemble. For example, the count of edges in
one instance should provide an estimate of the edge probability in the
ensemble.

This paper is structured as follows. In Section~\ref{sec:1} we
describe the theoretical framework of exponential families of random
graphs. Building on this, Section~\ref{sec:meas-struct-netw} contains
our main results. We interpret common observables on graphs in our
framework and shed new light on their
interpretation. Section~\ref{sec:netw-struct-simple} contains concrete
examples of our models with few parameters and discusses special
features of sampling procedures. Following the discussion in
Section~\ref{sec:discussion} is an Appendix which contains technical
details.

\section{Basic setting --- exponential families of random graphs}
\label{sec:1}
We consider undirected random graphs without self connections,
specified by an adjacency matrix $A$. Denote by $N$ the number of
nodes, $[N] \defas \set{1,\ldots,N}$ the set of nodes, and $E \defas
\set{(i,j) \in [N]\times [N] : i \neq j}$ the set of off-diagonal
indices of $A$. In a given graph each edge is either present or not,
therefore we are dealing with $\abs{E}$ binary random
variables. Denote by $\mathcal{X} \defas \set{0,1}^{E}$ the set of
states of this collection of random variables. For any subset $B$ of
potential edges we denote $\mathcal{X}_{B} \defas \set{0,1}^{B}$. 

In this labeled setting the probability of a random graph is given by
the probability of its adjacency matrix $A = (a_{e})_{e \in E}$. 
A random graph $G$ is described by binary random variables with state
space $\mathcal{X}$ and a probability
\[
P(G)=P(a_{e_1},\ldots,a_{e_{N (N-1)}})  \;.
\] 
A distribution $P(G)$ is often called a \emph{graph ensemble} in
the following. We consider so called
exponential families, classes of graph ensembles with a
particularly nice structure and interpretation. 
When used to describe probability distributions
for random graphs they have been termed ``exponential random
graphs'', ``$p^{*}$ models'', or ``logit-models'' in the
literature. For a recent overview see~\cite{Robins2007}. Here, we
utilize families $\mathcal{E}_k$ that consist of the
distributions with interactions between at most $k$ units.  Let $f$ be
a function mapping states $(a_{e_1},\ldots,a_{e_{N (N-1)}})$ to the
reals. With the usual addition and multiplication by real numbers,
these functions form a vector space
\[
\mathbb{R}^{\mathcal{X}} \defas \set{f : \mathcal{X} \to \mathbb{R}}.
\]
Of course, any (real-valued) observable is such a function. Our
systematic approach to quantifying structure consists in considering
natural bases of this space. We can then express well known
observables in terms of these bases, yielding a better understanding
of the relation between observables.

The probability measures
\begin{equation*}
  \mathcal{P}(\mathcal{X}) \defas \set{ P \in \mathbb{R}^{\mathcal{X}} : P(x) > 0,
    \sum_{x\in\mathcal{X}} P(x) = 1}\;,
\end{equation*}
form a subset of $\mathbb{R}^{\mathcal{X}}$, which has the geometry of
a simplex. Its closure, where $P(x)=0$ is allowed, is denoted by 
$\overline{\mathcal{P}}(\mathcal{X})$. The exponential map assigns to
each function a strictly positive probability measure:
\[
  \exp : \mathbb{R}^{\mathcal{X}} \to \mathcal{P}(\mathcal{X}) 
  \qquad f \mapsto \frac{\exp(f)}{\sum_{x\in\mathcal{X}} \exp(f(x))}\;.
\] 
Here, $\exp(f)$ is to be taken coordinatewise. Using the
exponential map, there is a natural way to define exponential families
of probability measures by considering families $\mathcal{E}$
that are exponential images of linear subspaces of $\mathbb{R}^{\mathcal{X}}$. 
One natural class of such subspaces is given by limiting the
interaction order. Following~\cite{Ay2006a,Kahle2009} one can define
\begin{equation*}
  {\mathcal I}_B \defas \left\{ f \in {\mathbb R}^\mathcal{X} : 
    f(x_B,x_{E \setminus B}) = f(x_B, x_{E
      \setminus B}') \mbox{ for all $x_B \in \mathcal{X}_B$, $x_{E \setminus B}, 
      x_{E \setminus B}' \in \mathcal{X}_{E
        \setminus B}$} \right\}\;,
\end{equation*}
as the space of functions depending only on the subset $B\subset E$ of their
arguments. Then \[\mathcal{I}_{k} = \sum_{\abs{B} \leq k}
\mathcal{I}_{B}\] is the space of functions depending on at most $k$
of their arguments. Here, the sum over $\mathcal{I}_{B}$ is to be
understood as their span inside $\mathbb{R}^{\mathcal{X}}$. This
definition leads to a hierarchy of exponential families
\begin{equation}
\label{eq:amari-hierarchy}
\mathcal{E}_{1} \subsetneq \ldots \subsetneq \mathcal{E}_{N(N-1)-1} 
\subsetneq \mathcal{P}(\mathcal{X})\;,
\end{equation}
which is studied in information geometry~\cite{Amari2001}. It 
allows to model networks by
considering interactions of successively increasing order between
their parts. It has been used to quantify the amount and degree of
interaction in dynamical systems in
a systematic fashion in~\cite{Kahle2009}. 

The notion of interaction order can be understood from the fact that
any $P \in \mathcal{E}_k$ has a (non-unique) representation as
\[
P(G) = \prod_{B\subset E : \abs{B} = k} \Phi(x_B) = \frac{1}{Z}\exp
\left( \sum_{B\subset E: \abs{B} = k} \phi_B(x_B) \right) \;.
\]
Thus, $P \in \mathcal{E}_k$ means that 
\begin{align}
P(G)=&\frac{1}{Z} e^{-\Ha(G)}  \notag \\
\Ha(G)=&\sum_{e} c_e f_{e}(a_e)+\sum_{e_1,e_2} c_{e_1,e_2}
f_{e_1,e_2}(a_{e_1},a_{e_2})+\ldots  \label{eq:3} \\ &
\ldots+\sum_{e_1,e_2,\ldots,e_k} c_{e_1,e_2,\ldots,e_k}
f_{e_1,e_2,\ldots,e_k}(a_{e_1},a_{e_2}, \ldots ,a_{e_k})\notag  \;.  
\end{align}

As mentioned above \emph{any} collection
$\mathcal{O}_{1},\ldots,\mathcal{O}_{s}$ of observables defines a
linear space as their span \[\mathcal{L} = \set{f \in
  \mathbb{R}^{\mathcal{X}} : f = \sum_{i=1}^{s} c_{i}
  \mathcal{O}_{i},\; c_{1},\ldots,c_{s} \in\mathbb{R}}\;,\] which in
turn defines an exponential family $\mathcal{E}_{\mathcal{L}} \defas \exp
(\mathcal{L})$. This exponential family is the collection of maximum
entropy distributions for fixed expectation values of the
observables. In particular, the mean values of the observables in a
sample form a sufficient statistics for the model
$\mathcal{E}_{\mathcal{L}}$. Given data, one can determine the mean
values of the observables on the data and then find a \emph{unique}
$P$ in the closure $\overline{\mathcal{E}}_{\mathcal{L}}$ which has
the same statistics as the data and maximal entropy among all such
distributions. Finding this estimate in practice can be
computationally expansive for a general linear space. In practice, an
algorithm called \emph{iterative proportional fitting} is used
\cite{deming40}. It is implemented in \textsc{cipi}~\cite{cipi} and
statistical software packages like \textsc{loglin} inside the
software $\textsc{R}$. Since this method works directly on the vectors
in $\mathbb{R}^{\mathcal{X}}$ it is limited to small
$\mathcal{X}$. Less than $N=20$ elements can be feasible. It is well known
that for $\mathcal{E}_{1}$ the maximum entropy distribution is just
the product of the one-dimensional marginals of the data.

\subsection{Undirected Graphs and subgraph counts}
\label{sec:undir-graphs-subgr}
In the following we specialize the general theory to the case of
undirected unlabeled homogeneous graphs. Here $E = \set{(i,j) \in
  [N]\times [N] : i<j}$ is the set of potential edges, resulting
in a symmetric adjacency matrix $A = (a_{ij})_{i,j = 1,\ldots,N}$, by
setting $a_{ij} = a_{e}$ if $(i,j) \in E$ or $(j,i) \in E$, and
$a_{ii} = 0$. 

Unlabeled graphs are defined as equivalence classes and working with
them in practice becomes infeasible quickly. It is a curiosity of
complexity theory that the graph isomorphism problem is in the class
NP, but neither known to lie in P, nor to be NP-complete. In any case,
at the current time there is no fast algorithm to determine whether
two unlabeled graphs are isomorphic. Due to this unavoidable
restriction we will always work with adjacency matrices. In particular
the partition function is a sum over symmetric adjacency matrices and
$P(H)$, for some unlabeled graph $H$ denotes the sum of probabilities
of adjacency matrices which have $H$ as their unlabeled graph. This is
complemented by the \emph{homogeneity} that we assume in this
setting. It means that we consider observables that are only evaluated
on a small part of the system, and the value should show no systematic
differences when varying the position. In particular, the probability
of finding an edge should be the same for any pair of
vertices. 
In the setting that was described in
Section~\ref{sec:1} this homogeneity is a certain symmetry
requirement. For instance the \ER graphs emerge from the exponential
family where the linear space is the one-dimensional span of the edge
count observable.

Generalizing from \ER graphs to higher dependency structures, natural
observables are the subgraph counts, defined as follows: Given a graph
$G$ with potential edge set $E$, we define the subgraph counts of a
subgraph $H$ as
\begin{equation}
  \label{eq:subgraphcountsdef}
  n_{H} \defas \#\set{\text{unlabeled subgraphs $H$ of $G$}}\;.
\end{equation}
For example, we denote by $n_{-}$ the number of edges and by
$n_{\nine}$ the number of triangles. For any undirected graph $G$, the
counts of subgraphs with at most $k$ edges form a sufficient
statistics for the exponential family $\mathcal{E}_{k}$ considered
above. This can be seen as follows. For each set $B\subset E$ of
potential edges, we can define a function which takes the value one
precisely if all edges in $B$ are present, \ie it counts the subgraph
defined by the labeled graph $B$\footnote{The graph corresponding to
  $B \subset E$ specifies the relations between the edges in $B$. As
  an example consider $B_{\two} = \{(1,2), (1,6)\}$ where the two
  edges share vertex $1$ compared to $B_{\one} = \{(1,2), (4,7)\}$
  where they are disconnected.}:
\begin{equation}
  \label{eq:5}
f_{B}(G):=\prod_{e \in B} a_e  \;.
\end{equation}
A classical argument from the theory of Markov random fields
\cite{winkler_image_analysis} shows that these functions $\set{f_{B} :
  B\subset E}$ form a basis of the whole space
$\mathbb{R}^{\mathcal{X}}$, while with $k = \abs{B}$ we have $f_{B}
\in \mathcal{I}_{k}$. Uniqueness of the coefficients with
respect to this basis depends on a choice of a reference
configuration, the so called \emph{vacuum} state. In our case the vacuum is
the empty graph. The statement about sufficiency follows if one
observes that the homogeneity requirement translates into a condition
on the coefficients in \eqref{eq:3}: Coefficients
$c_{e_{1},\ldots,e_{l}}$ for different $e_{1},\ldots,e_{l}$
representing the same undirected unlabeled subgraph are required to be
the same. Therefore the counts of unlabeled subgraphs with $k$ edges
(which are just sums over the $f_{B}$ for all $B$ representing a
specific subgraph) span the linear subspace space of functions
depending only on $k$ of their arguments, and taking equal values
whenever these $k$ arguments represent the same subgraph. Summarizing
we have
\begin{equation}
  \label{eq:1}
  \sum_{B \sim H} f_{B} = n_{H}\;,
\end{equation}
where the summation runs over all sets $B$ which define a subgraph
isomorphic to an unlabeled graph $H$. 

Note that fixing the number of
vertices breaks the relation with the hierarchy in
\eqref{eq:amari-hierarchy}. Consider as an example the full model with
subgraph counts up to 4 vertices. The energy has the form
\begin{equation}
  \Ha = c_{-}n_{-} + c_{\two}n_{\two} + c_{\nine}n_{\nine}  + \ldots
  + c_{\six}n_{\six} + c_{\seven}n_{\seven}\;.
\end{equation}
This distribution is an element of $\mathcal{E}_{6}$, but not all
elements of $\mathcal{E}_{6}$ are of this form, as we have not used a
subgraph count for 6 edges forming a chain, which would be a subgraph
on 7 vertices. It is also important to notice that changing one of
the coefficients will generally change all of the expected counts.

Apart from the subgraph counts, we often use the subgraph
probabilities $p_{H}$, that is, the probabilities for observing the
subgraph $H$ when drawing a random graph from the ensemble $P$ on
randomly selected nodes. This can be written as an expectation
value\footnote{We use the notation $\langle \cdot \rangle$ for
  expectation values with respect to the graph ensemble $P$.} with
respect to the distribution $P$ as
\begin{equation}
  \label{eq:2}
  p_{H} \defas \left\langle f_{B} \right\rangle, \qquad B \sim H\;,
\end{equation}
where $B \subset E$ is any set of edges whose unlabeled graph is
$H$. That any such $B$ can be chosen is a consequence of the
homogeneity that we require for our model. If the subgraphs whose
counts we use as observables are small enough (when compared to the
size of the network $G$), the homogeneity assumption allows to use
counts on a single given network to estimate the ensemble expectation
values. Quantities derived from a single network are denoted by a
hat as in $\widehat{n}_{H}$.

 When a subgraph probability $p_{H}$ is
estimated from a single network, it is given by the count of that
subgraph, normalized by the maximum possible number of occurrences of
that subgraph. This in turn is just the number of occurrences of $H$
in the fully connected network $F$:
\begin{equation}
  \widehat {p}_{H} = \frac{n_{H}(G)}{n_{H}(F)}\;.
\label{p_H}
\end{equation}

\section{Measures of structural network properties}
\label{sec:meas-struct-netw}
\subsection{Link density}
One of our main aims is to find good sets of network observables that
capture the important structural properties of a graph. Obviously the
first property is the number of links $n_{-}$ or the link density 
\begin{eqnarray}
  \label{eq:6}
  \widehat{p}_{-}&=&\frac{n_{-}(G)}{n_{-}(F)} \\
  &=&\frac{2 n_{-}}{N (N-1)} \;,
\end{eqnarray}
with $F$ denoting the fully connected graph. \\
If only the expectation value of the link density or the number of
links is specified, the corresponding maximum entropy ensemble is the
ensemble of \ER random graphs or Bernoulli graphs. Its Hamiltonian is
simply \be \Ha(G)=c_{-} n_{-}= c_{-} \sum_{(i,j) : i<j} a_{ij} \;.  \ee
The main property of this ensemble is that there are no statistical
dependencies between the links. The degrees of the nodes are
distributed according to a Bernoulli distribution fully determined by
the mean node degree.

\subsection{Degree distribution}

A distribution different from the Bernoulli distribution introduces
statistical dependencies between the potential links. The resulting
random graph ensemble will therefore be different from the Bernoulli
graphs. How does the degree enter our framework? In a first step one
might label the nodes using some labeling $\pi$, and assign to each
node an expected degree. This leads to exponential
random graph model \be \Ha(G,\pi)=\sum_i c_i \sum_{j} a_{\pi(i) \pi(j)},
\ee with the $c_i$ determining the degree of node $\pi(i)$. 
In a second step one considers an ensemble
of ensembles of random graphs, where the different ensembles
correspond to different labellings, i.e. permutations of the node
degrees. This leads to a probability distribution that is a convex
combination of distributions $P \in \mathcal{E}_1$, and thus generally
not contained in $\mathcal{E}_1$ but in a so called mixture model. Moreover, this approach
does not lead to a parameterization of the form (\ref{eq:3}).

Our next aim is to understand how the degree distribution can be
specified within the exponential families $\mathcal{E}_k$. 
The key point is that the
$k$-star\footnote{A $k$-star is a subgraph consisting of one central
  node that is connected to $k$ other peripheral nodes, i.e. it
  contains $k$ links.} counts allow to determine the moments of the
degree distribution. With $d_i=\sum_{j} a_{ij}$ being the degree of
node $i$ we call $P(d)$ the probability that a randomly chosen node
has degree $d$. We find \bes P(d)&=&\sum_{G} P(G)
\frac{1}{N} \sum_{i=1}^N \delta_{d_i(G),d} \;.  \ees The moments of
the degree distribution are \be \mu_k=\langle d^k \rangle =
\sum_{d=0}^\infty d^k P(d) \;. \ee For a given graph, we also have the
empirical degree distribution
\[
\widehat{P}_G(d)=\frac{1}{N} \sum_{i=1}^N \delta_{d_i(G),d} \;,
\]
with the moments
\be
\widehat{\mu}_k=\frac{1}{N} \sum_{i=1}^N d_i^k \;.
\label{moments}
\ee
Note that $\langle \widehat{\mu}_k \rangle =\mu_k $. There is a direct 
relationship between the moments of the degree distribution and the numbers of $k$-stars
\be
n_k=\sum_{i=1}^N {d_i \choose k}  \;.
\label{n-k-star}
\ee 
Note that with this definition the number of $1$-stars is two times the numbers of links $n_1=2
n_{-}$. Thus looking for the maximum entropy distribution for graphs
with given moments of the degree distribution up to order $k_{\max}$
corresponds to the exponential random graph model having
non-zero coefficients only for $k$-stars with $k \le k_{\max}$. In
particular, this distribution lies in $\mathcal{E}_{k_{\max}}$. The
parameterization using the $k$-stars on the one hand side \be 
\Ha=\sum_{k=1}^{k_{\max}} c_k n_k\;, \ee and the moments of the empirical
degree distribution of the graph \be \Ha=\sum_{k=1}^{k_{\max}}
c^{(d)}_k \widehat{\mu}_k\;,
\label{H_moments}
\ee 
can be converted into each other using
\bes
n_k &=& \frac{1}{k!} \sum_{i=1}^N \sum_{m=1}^k s(k,m) d_i^m \\
&=& \frac{N}{k!} \sum_{m=1}^k s(k,m) \widehat{\mu}_m \;. 
\label{n-k-star2}
\ees with $s(k,m)$ being the Stirling numbers of the first kind. The
inverse relationship between the $k$-star counts and the moments of
the degree distribution can be expressed using the Stirling number of
the second kind $S(k,m)$: \be \widehat{\mu}_k=\sum_{m=1}^k S(k,m)
\frac{k!}{N} n_m \;.  \ee For the parameters this leads to the
relationships \bes
c_m&=& \sum_{k=m}^{k_{\max}} \frac{k!}{N} S(k,m) c^{(d)}_k \\
c^{(d)}_m&=& \sum_{k=m}^{k_{\max}} \frac{N}{k!} s(k,m) c_k \;.  \ees
The parameterization (\ref{H_moments}) might still not be the best way
to explore different degree distributions, because of the dependencies
between the different moments. Instead of using the empirical moments
(\ref{moments}) as observables, one could think of observables that
can be independently varied more easily, such as the variance,
skewness, and kurtosis. Let us look in more detail at the
variance, the other cases are similar.
In the \emph{two star model} \bes \Ha(G)&=&c_{-} n_{-}+c_{\two} n_{\two}
\label{H_two_star}\\
&=&c_1 n_1+c_2 n_2 \;.
\ees 
with $c_1=c_{-}/2$ and $c_2=c_{\two}$, the resulting probability
distribution can be equivalently parameterized by the pairs
$(c_1,c_2)$, $(\mu_1,\mu_2)$ or $(\mu_1,\mu_2-\mu_1^2)$, the last
being the mean and the variance of the degree distribution. 
The variance of the empirical degree distribution of a graph $G$ is 
\be
\mbox{var}(\widehat{P}(d))=\widehat{\mu}_2-{\widehat{\mu}_1}^2 \;.
\ee
One might think of a Hamiltonian of the form
\be
\Ha(G)=c^{(d)}_1 \widehat{\mu}_1+\widetilde{c}^{(d)}_2 \left( \widehat{\mu}_2-{\widehat{\mu}_1}^2 \right) \;.
\label{H_variance}
\ee This model is different from the two star model because
it involves a non-linear transformation of the observables. 
\subsection{Cluster coefficient}
The cluster coefficient can be defined as three times the ratio between 
the numbers of triangles and the numbers of two stars in a given graph:
\bes
\widehat{C}&=&\frac{3 n_{\nine}}{n_{\two}} \\
&=&\frac{\widehat{p}_{\nine}}{\widehat{p}_{\two}}\;,
\ees
with
\bes
\widehat{p}_{\two}&=&\frac{2 n_{\two}}{N (N-1) (N-2)} \\
\widehat{p}_{\nine}&=&\frac{6 n_{\nine}}{N (N-1) (N-2)} \;.
\ees
Thus the cluster coefficient for the ensemble $P$ measures the
conditional probability that if for three randomly selected nodes 
 one node is connected to the two others, these are also connected.
\be
C=p(a_{ij}=1|a_{ik}=1,a_{jk}=1) \;.
\ee
In the context of social networks this property is also called ``transitivity'' because it means the probability that the friend of my friend is also my friend. If there are no statistical dependencies between the links, we would expect
\be
C_{\text{ind}}:=p_{-} \;.
\ee
Moreover, if there are statistical dependencies only between pairs of links ($P(G) \in \mathcal{E}_2$), such as in the two star model, one might expect
\[
p(a_{ij}=1|a_{ik}=1,a_{jk}=1)=p(a_{ij}=1|a_{ik}=1)\;,
\]
or
\be
\frac{p_{\nine}}{p_{\two}}=\frac{p_{\two}}{p_{-}} ,
\ee
respectively, and therefore the cluster coefficient would be equal to
\be
\frac{p_{\two}}{p_{-}} \;.
\label{C^{(2)}}
\ee
This is, however, not the correct expression for the two star
model. Already the case of only three nodes provides an example:
\[
P(G)=P(a_{1,2},a_{2,3},a_{3,1})=\frac{1}{Z} \exp(c_{-} n_{-}+c_{\two} n_{\two}) \;.
\]
There we have
\besn
Z P(0,0,0)&=&1 \\
Z P(1,0,0)&=&Z P(0,1,0)=Z P(0,0,1)=h_1=\exp c_{-}  \\
Z P(1,1,0)&=&Z P(1,0,1)=Z P(0,1,1)=h_2=\exp (2 c_{-}+c_{\two})  \\
Z P(1,1,1)&=&h_3=\exp (3 c_{-}+3 c_{\two})=p_{\nine}  \\
Z&=&1+ 3 h_1+ 3 h_2+h_3  \;.
\eesn
Hence, the cluster coefficient is
\besn
C&=&\frac{p_{\nine}}{p_{\two}}=\frac{P(1,1,1)}{P(1,1)}=\frac{P(1,1,1)}{P(1,1,1)+P(1,1,0)} \\
&=&\frac{h_3}{h_2+h_3}=\frac{1}{1+h_2/h_3} \;.
\eesn 
On the other hand (\ref{C^{(2)}}) becomes
\besn
\frac{p_{\two}}{p_{-}}&=&\frac{h2+h3}{h2+h3+h2+h1} \\
&=&\frac{1}{1+h_2/h_3 (\frac{h_1/h_2+1}{h_2/h_3+1})} \neq C  \;.
\eesn
If the three random variables are only a subset of a larger set of random variables as in the case of larger networks, things become even more complicated. \\
Nevertheless, if $n_{-}$ and $n_{\two}$ are sufficient statistics for
the two star model, we should be able to express the cluster
coefficient by these two variables. In particular, we should be able
to express the expected number of triangles by the expected number of
two stars and the expected number of links.
\subsection{Markov graphs}
If the Hamiltonian contains only the numbers of $k$-stars and
triangles, it defines the so called Markov graphs
\cite{Frank1986}. This class of random graphs is well known in the
social network community. ``Markov'' here refers to the fact that in
these graphs the occurrence of links without a common node is
statistically independent. The only subgraphs where all pairs of links
have a common node are the $k$-stars and the triangles. From what we
have discussed so far it becomes clear that these models can account
already for a large range of degree distributions in contrast to the
statement sometimes found in the literature that the exponential random
graph models of the social network community only accounts for
Poissonian degree distributions~\cite{Boerner2007}.
\subsection{Assortativity}
Another widely studied property of a graph is its assortativity or
disassortativity. In an assortative graph, high degree nodes are
prevalently connected to other high degree nodes and low degree nodes
to low degree ones. In disassortative graphs, high degree nodes tend
to be connected to low degree nodes. A simple way to measure this
property is the correlation coefficient between the remaining degrees
of two connected nodes~\cite{Newman2002}. ``Remaining'' degree refers
to the degree of a node after subtracting one for the link connecting
this node to the other one. Empirical investigations showed that most
social networks are assortative, while the Internet or biological
networks are rather disassortative~\cite{Newman2002}.

For an edge
$a_{ij} = 1$ the remaining degrees at either side of the edge are given by
\begin{eqnarray*}
d^{r,i}_{ij} & = & \sum_{k \neq i,j} a_{ki} = d_i - 1\;, \\
d^{r,j}_{ij} & = & \sum_{l \neq i,j} a_{jl} = d_j - 1\;.
\end{eqnarray*}
The assortativity is then given by the correlation coefficient between the remaining degrees
at either side of an edge:
\[ r^2 = \frac{\langle (d^{r,i} - \langle d^{r,i} \rangle) (d^{r,j} - \langle d^{r,j} \rangle) \rangle}{
\sqrt{\langle (d^{r,i} - \langle d^{r,i} \rangle)^2 \rangle
\langle (d^{r,j} - \langle d^{r,j} \rangle)^2 \rangle }}\;. \]
Using that we consider undirected graphs, i.e. $A$ is symmetric, $\langle (d^{r,i})^n \rangle = \langle (d^{r,j})^n \rangle$ 
for $n = 1,2,\ldots$, and linearity of expectation values, this simplifies to
\begin{equation} 
r^2 = \frac{\langle d^{r,i} d^{r,j} \rangle - \langle d^{r,i} \rangle \langle d^{r,i} \rangle}{
\langle (d^{r,i})^2 \rangle - \langle d^{r,i} \rangle^2}\;. \label{equ:r2}
\end{equation}

All relevant quantities can now be expressed in terms of subgraph counts (see appendix for details):
\be
\widehat{r^2} = \frac{\frac{n_{-}}{n_{\two}} \left(\frac{3 n_{\nine}}{n_{\two}}+\frac{n_{\three}}{n_{\two}} \right)-1}{\frac{n_{-}}{n_{\two}} \left(\frac{3 n_{\four}}{n_{\two}}+1 \right)-1}\;.
\ee
In order to express the assortativity by the subgraph probabilities we list again all subgraph probabilities including the missing ones:
\bes
\widehat{p}_{-}&=&\frac{2 n_{-}}{N (N-1)} \\
\widehat{p}_{\two}&=&\frac{2 n_{\two}}{N (N-1) (N-2)} \\
\widehat{p}_{\nine}&=&\frac{6 n_{\nine}}{N (N-1) (N-2)} \\
\widehat{p}_{\four}&=&\frac{6 n_{\four}}{N (N-1) (N-2) (N-3)} \\
\widehat{p}_{\three}&=&\frac{2 n_{\three}}{N (N-1) (N-2) (N-3)}   
\ees
Thus
\bes
\widehat{r^2}&=&\frac{\frac{\widehat{p}_{-}}{(N-2) \widehat{p}_{\two}} \left(\frac{\widehat{p}_{\nine}}{\widehat{p}_{\two}}+\frac{(N-3) \widehat{p}_{\three}}{\widehat{p}_{\two}} \right)-1}{\frac{\widehat{p}_{-}}{(N-2) \widehat{p}_{\two}} \left(\frac{(N-3) \widehat{p}_{\four}}{\widehat{p}_{\two}}+1 \right)-1} \nonumber \\
&=& \frac{\left( \frac{\widehat{p}_{\nine}}{\widehat{p}_{\two}}-\frac{\widehat{p}_{\two}}{\widehat{p}_{-}}\right) +\left(\frac{\widehat{p}_{\three}}{\widehat{p}_{\two}} - \frac{\widehat{p}_{\two}}{\widehat{p}_{-}}\right) (N-3)}{\left( 1 -\frac{\widehat{p}_{\two}}{\widehat{p}_{-}}\right) +\left(\frac{\widehat{p}_{\four}}{\widehat{p}_{\two}} - \frac{\widehat{p}_{\two}}{\widehat{p}_{-}}\right) (N-3)}\;.
\ees
The assortativity coefficient is zero, if
\be
\frac{3 \cdot n_{\nine}}{n_{\two}} + \frac{n_{\three}}{n_{\two}}=\frac{n_{\two}}{n_{-}}\;,
\ee
or equivalently
\be
\frac{\widehat{p}_{\nine}}{\widehat{p}_{\two}}+\frac{(N-3) \widehat{p}_{\three}}{\widehat{p}_{\two}}=\frac{(N-2) \widehat{p}_{\two}}{\widehat{p}_{-}}\;,
\ee
which is a relation between conditional probabilities that is fulfilled in particular if
\begin{equation}
\frac{\widehat{p}_{\three}}{\widehat{p}_{\two}}=\frac{\widehat{p}_{\two}}{\widehat{p}_{-}} \label{cond_assort1}
\quad \text{ and } \quad 
\frac{\widehat{p}_{\nine}}{\widehat{p}_{\two}}=\frac{\widehat{p}_{\two}}{\widehat{p}_{-}} \;.
\end{equation}
Again, this does not mean that exponential random graphs with pairwise
interactions only, such as the two star model, have a vanishing
assortativity coefficient $r^2$. The same arguments as for the cluster
coefficient apply. For Markov graphs, defined as random graphs for
which links without a common node occur statistically independently,
we can make an interesting observation: Condition~(\ref{cond_assort1})
is fulfilled, and the assortativity is fully controlled by the cluster
coefficient.

\section{Network structure in simple exponential random graph models}
\label{sec:netw-struct-simple}
Let us consider an exponential random graph model
\begin{equation}
\Ha(G)=\sum_H c_H p_H(G)
\label{H_general}
\end{equation}
where the summation runs over some set of subgraphs. If we fix the
number of nodes, then (\ref{H_general}) defines an energy landscape in
the space of all graphs with $N$ nodes. High probability corresponds
to low energy, therefore the minima of (\ref{H_general}) should
correspond to the graphs that are most probable and therefore
``typical'' in the ensemble defined by this model.  A second possible
reason for a graph being typical is a high number of isomorphic
graphs, but for sufficiently low temperatures this effect will be
dominated by the effect of the energy.\footnote{We did not introduce a
  temperature explicitly, but it can be done easily be setting
  $\Ha(G)=E(G)/T$ with $E(G)$ being the energy and $T$ the
  temperature. Changing the temperature corresponds to a rescaling of
  all coefficients $c_H$ in $\Ha(G)$ by a constant factor.} Because we
expressed the energy using the subgraph probabilities (\ref{p_H}) it
is obvious that the empty graph has zero energy and the energy of the
fully connected graph is equal to the sum of all coefficients
$\Ha(F)=\sum_{H} c_{H}$. Thus we realize a first property of
(\ref{H_general}): If all coefficients $c_H$ are sufficiently
negative, the fully connected graph has minimal energy and is the most
probable and only typical graph. If, on the other hand, all
coefficients are sufficiently positive, the empty graph is the most
probable and therefore typical graph. We conclude that in order to get
non-trivial typical graphs, at least one coefficient has to have a
different sign then the other coefficients. A more detailed analysis
will show that additional requirements are needed in order to get
``interesting'' typical graphs. In the following we discuss this for
some simple exponential random graph models, and shed some light on
the difficulties that were reported by several authors that tried to
use them to describe real world networks~\cite{Handcock2003}.
\subsection{The two star model}
\begin{figure}
\begin{center}
\includegraphics[width=0.7\textwidth]{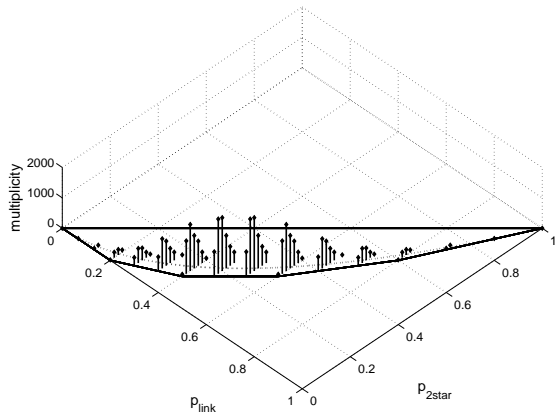}
\end{center}
\caption{Position of all graphs with $N=6$ nodes in the
  $(p_{-},p_{\protect \two})$-plane. On each point a bar is drawn
  according to its multiplicity, i.e.  the number of adjacency
  matrices with these link and two star counts. The dashed line shows
  the position of \ER graphs.}
\label{fig:2star}
\end{figure}
\begin{figure}
\begin{center}
\includegraphics[width=0.6\textwidth]{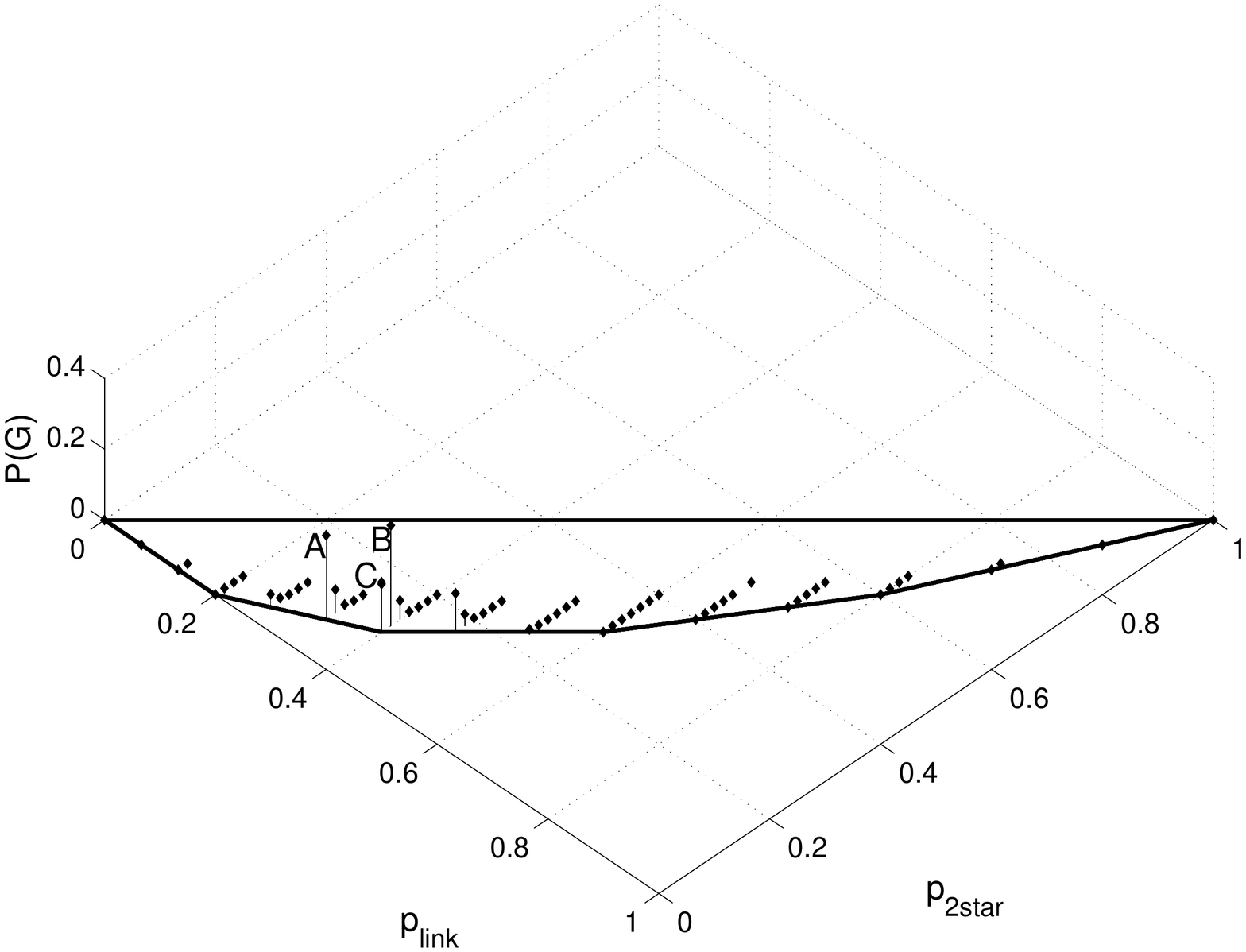}
\begin{minipage}[b]{0.3\textwidth}
\raisebox{8ex}{A}
\includegraphics[width=0.29\textwidth]{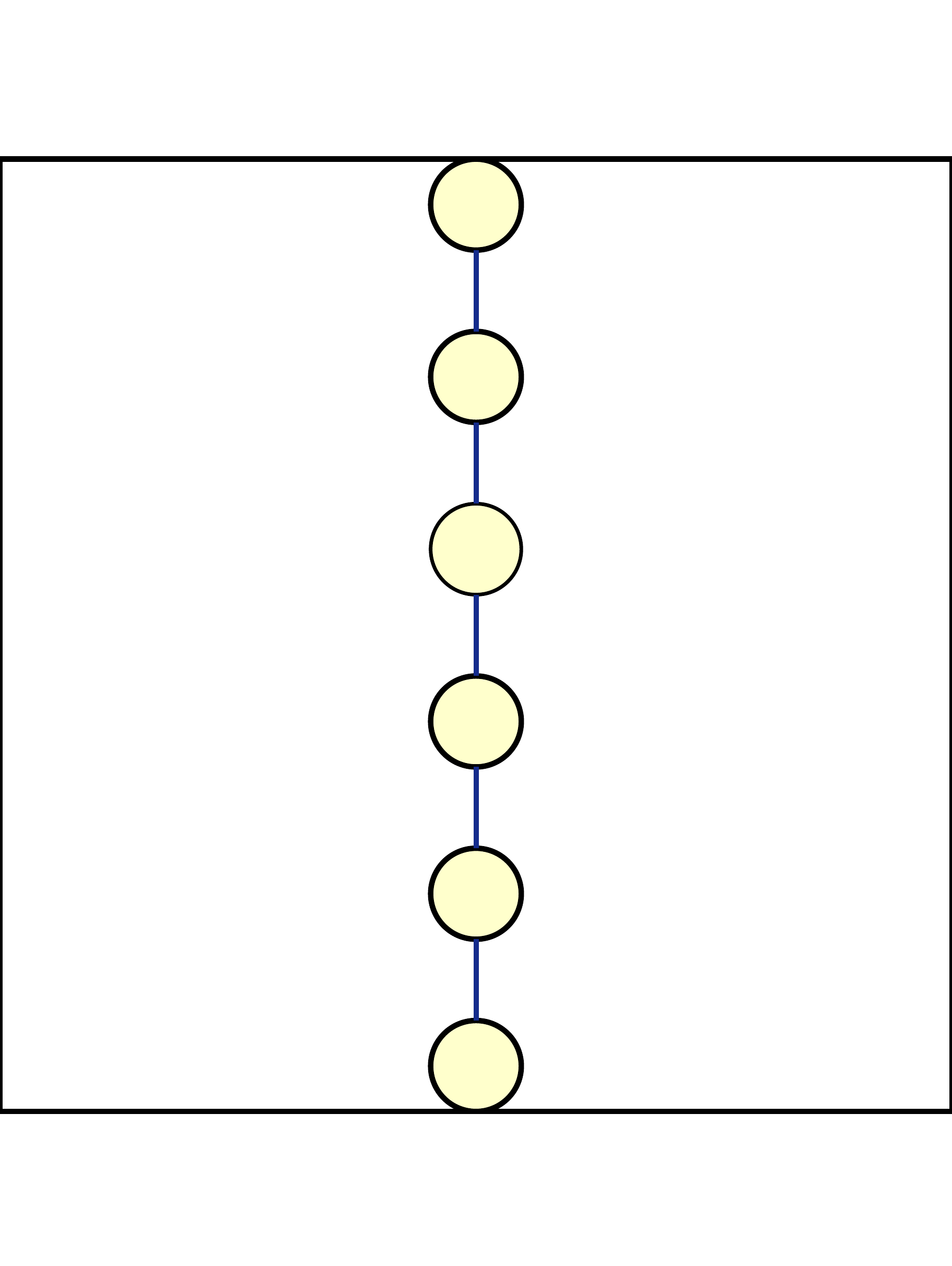}
\includegraphics[width=0.29\textwidth]{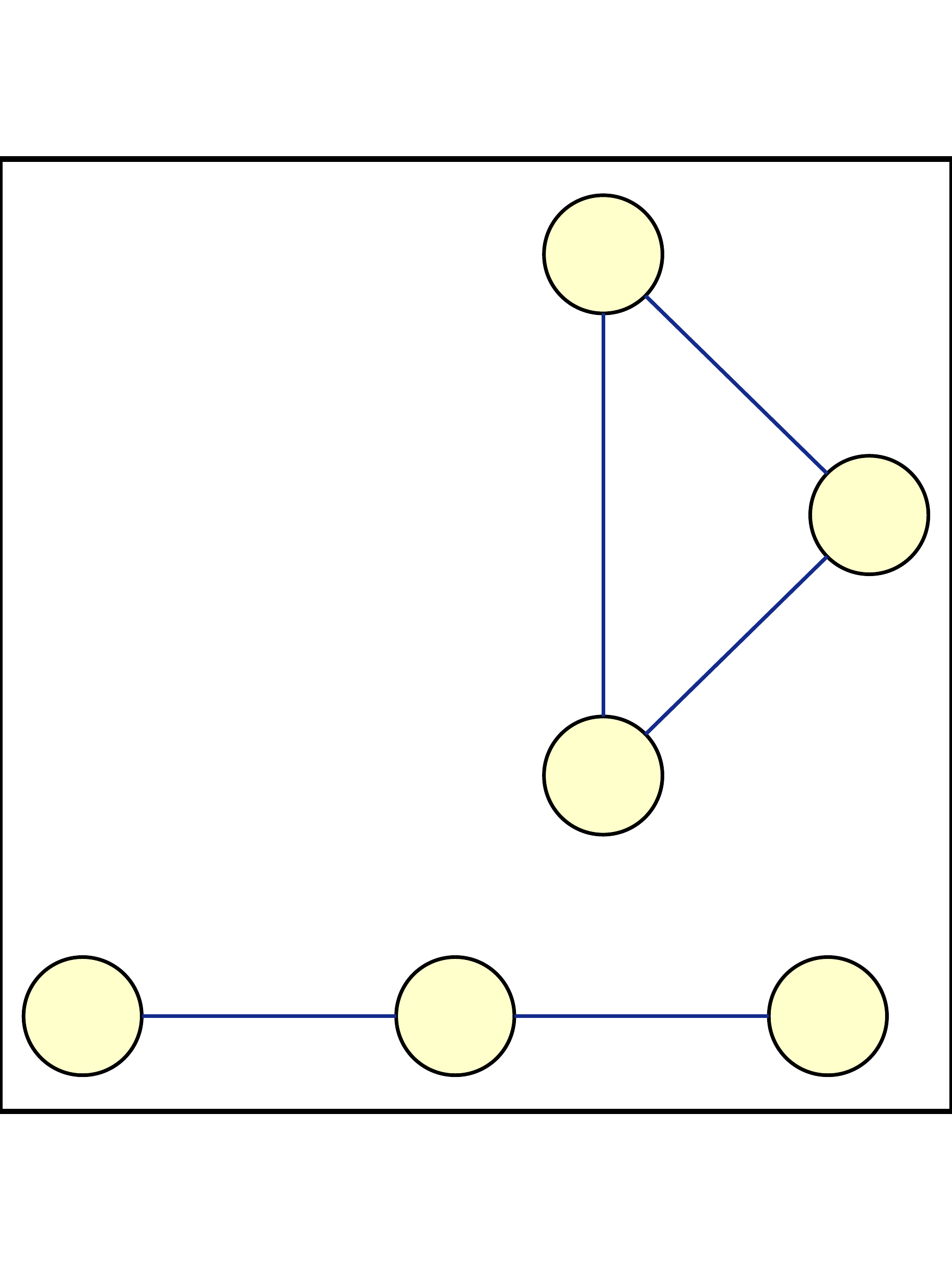}
\includegraphics[width=0.29\textwidth]{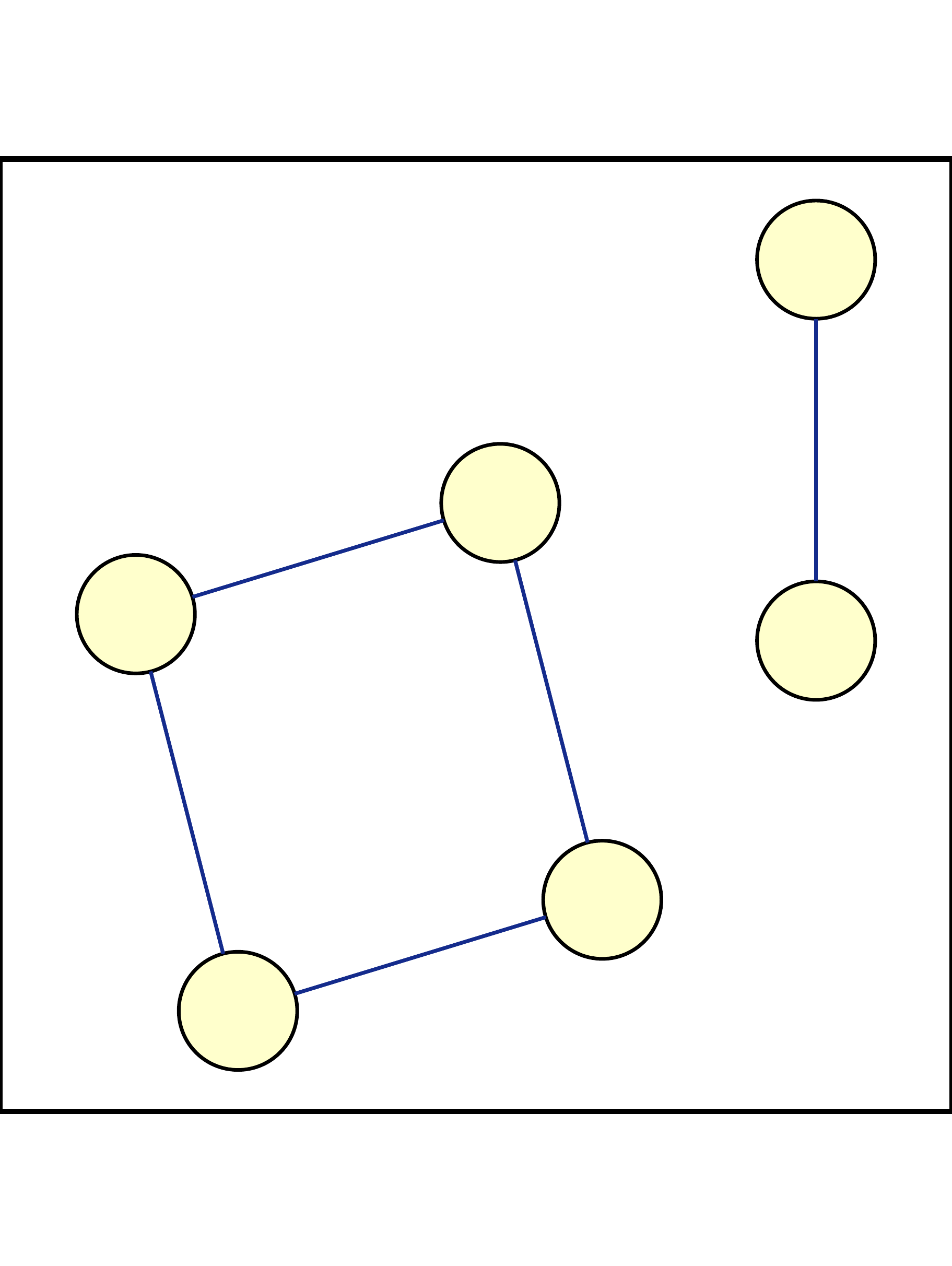} \\
\raisebox{8ex}{B}
\includegraphics[width=0.29\textwidth]{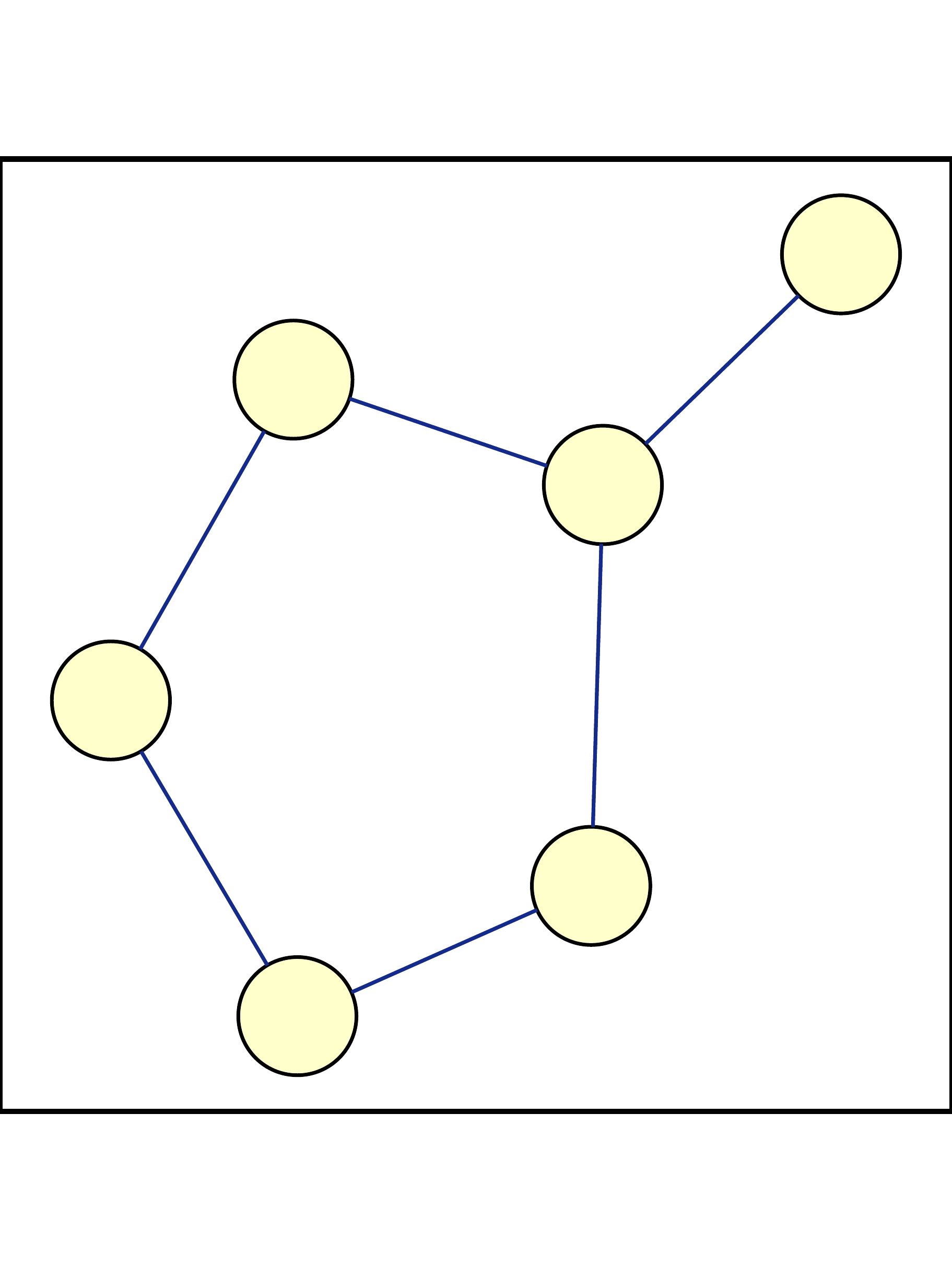}
\includegraphics[width=0.29\textwidth]{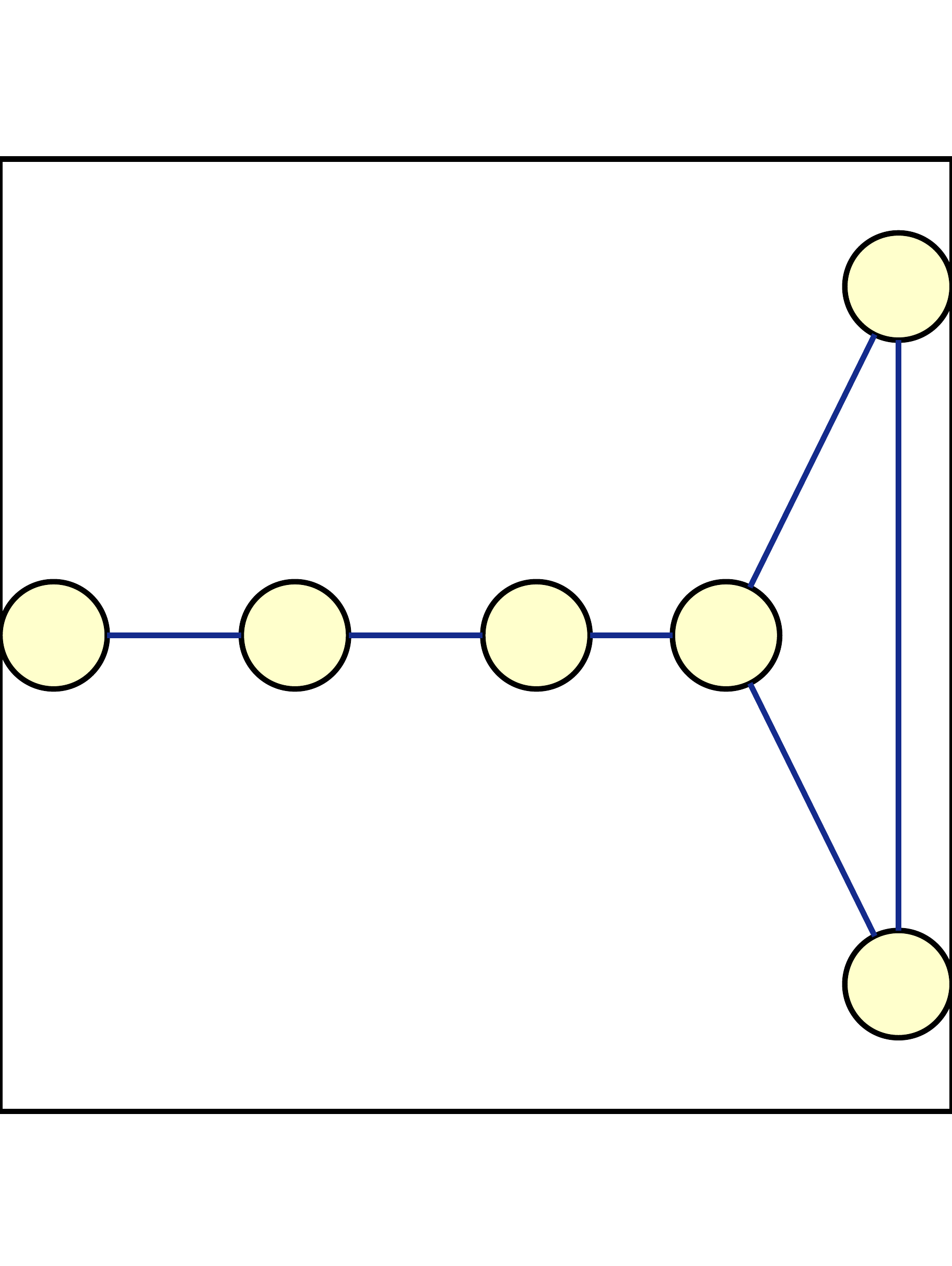}
\includegraphics[width=0.29\textwidth]{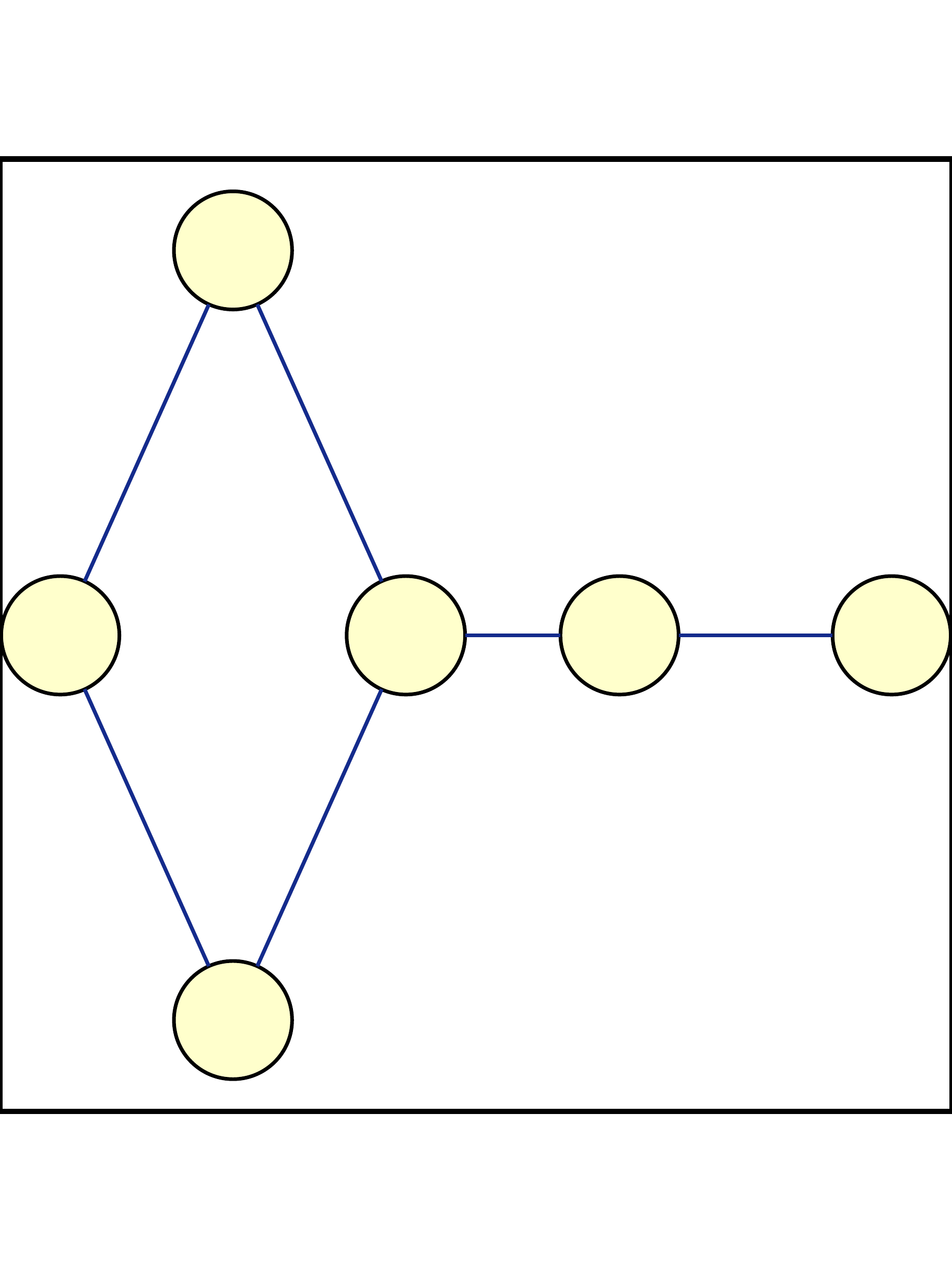}
\raisebox{8ex}{C}
\includegraphics[width=0.29\textwidth]{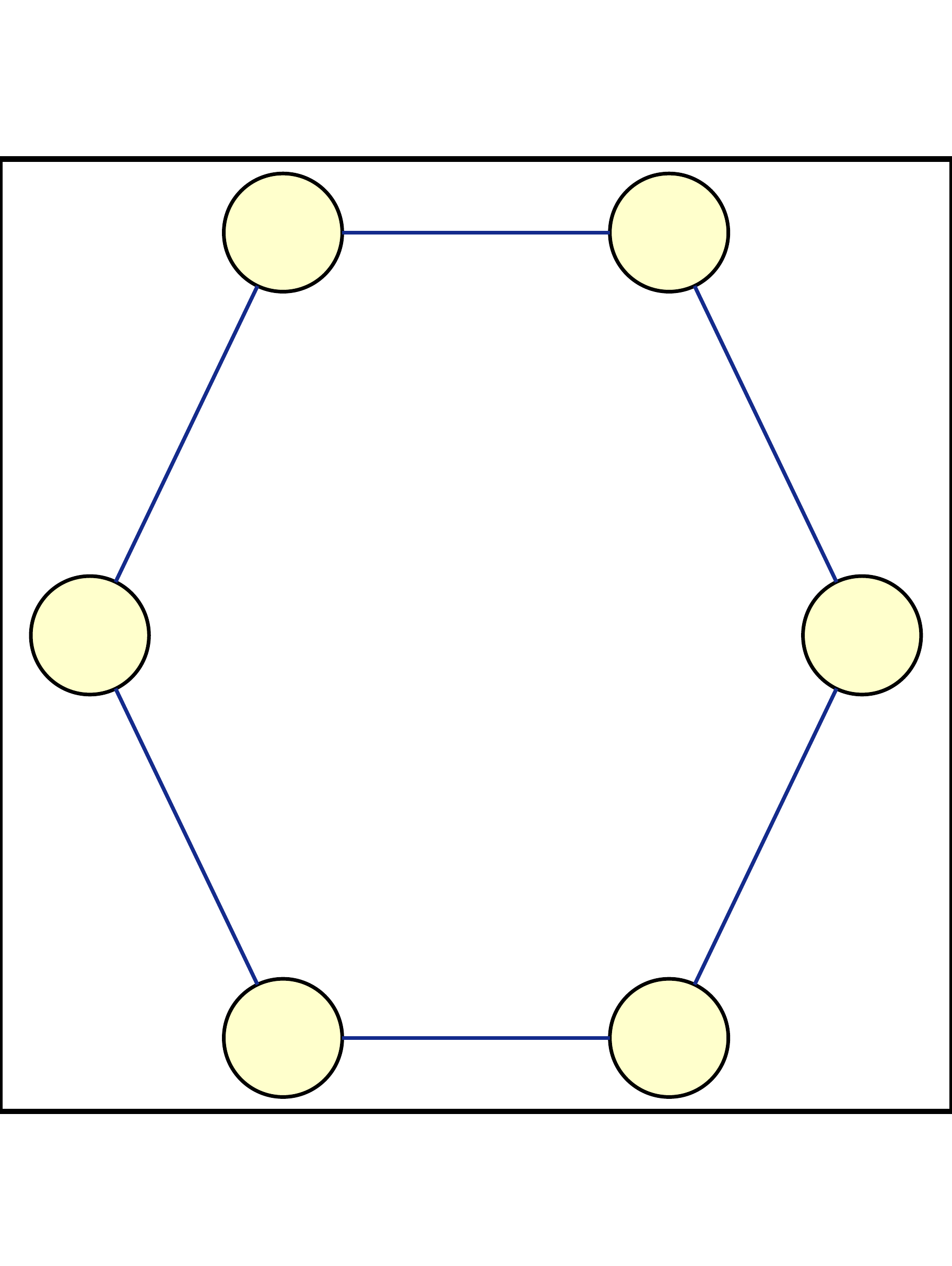}
\includegraphics[width=0.29\textwidth]{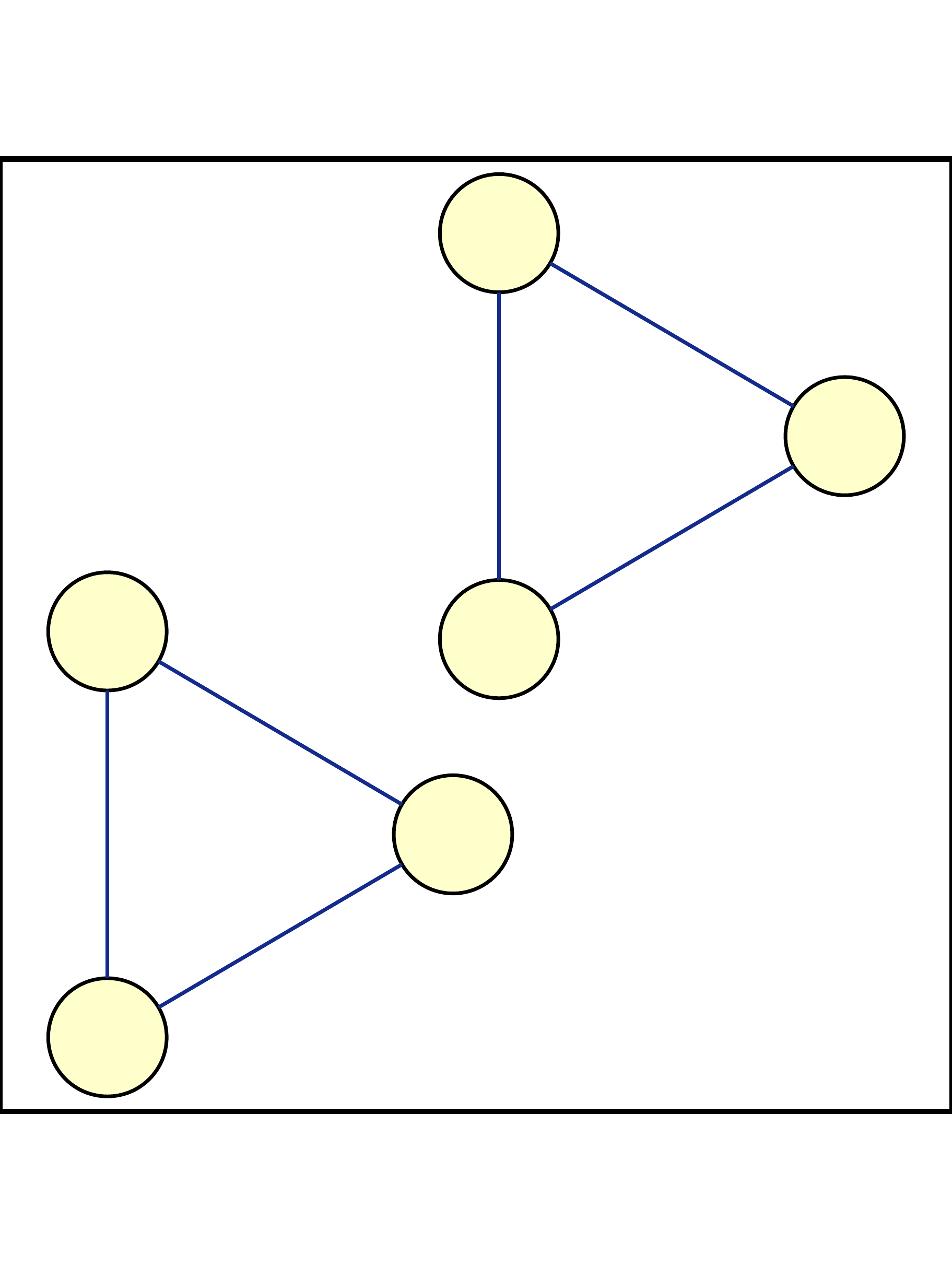}
\end{minipage}
\end{center}
\caption{Graph ensemble over graphs of six nodes as specified by the
  two star model with parameters $c_{-} = -80$ and $c_{\protect \two} = 120$.
  At the right typical graphs corresponding to the three most probable
  combinations of link and two star counts are shown.}
\label{fig:2star_ensemble}
\end{figure}

The two star model has the form
\be
\Ha=c_{-} p_{-}+c_{\two} p_{\two}  \;.
\label{two_star_model}
\ee Fig.~\ref{fig:2star} shows the position of all $6$-node graphs
in the $(p_{-},p_{\two})$-plane. The convex hull of these points 
defines all possible expectation values of $p_{-}$ and $p_{\two}$ for
two star models. By~linearity, the energy landscape is a plane in a
third dimension and extreme values lie on the boundary
of the convex region.
A positive value of $c_{-}$ and a negative value of
$c_{\two}$ result in the minimal energy graphs being located on
the upper boundary of the region. As visible in Fig.~\ref{fig:2star},
the empty and the fully connected graphs are the only 
graphs lying on this boundary. Therefore, for sufficiently low
temperatures, the graph ensemble is supported only on these two
graphs In the opposite case of negative $c_{-}$ and positive
$c_{\two}$ the minimal energy graphs lie on the lower boundary of the
region. These graphs have less two stars than the \ER graphs
with the same link density. This is the only structural property of
graphs that can be quantified by the two star model
(\ref{two_star_model}). Fig.~\ref{fig:2star_ensemble} gives an example of such a graph
ensemble and shows its typical graphs corresponding to the three most
probable combinations of link and two star counts. The two most probable
positions $A$ and $B$ have a higher energy than the graphs of lowest energy at $C$, 
but gain probability from their multiplicity (compare Fig.~\ref{fig:2star}).

\subsection{Cluster coefficient and assortativity}
\begin{figure}
\begin{center}
\includegraphics[width=0.7\textwidth]{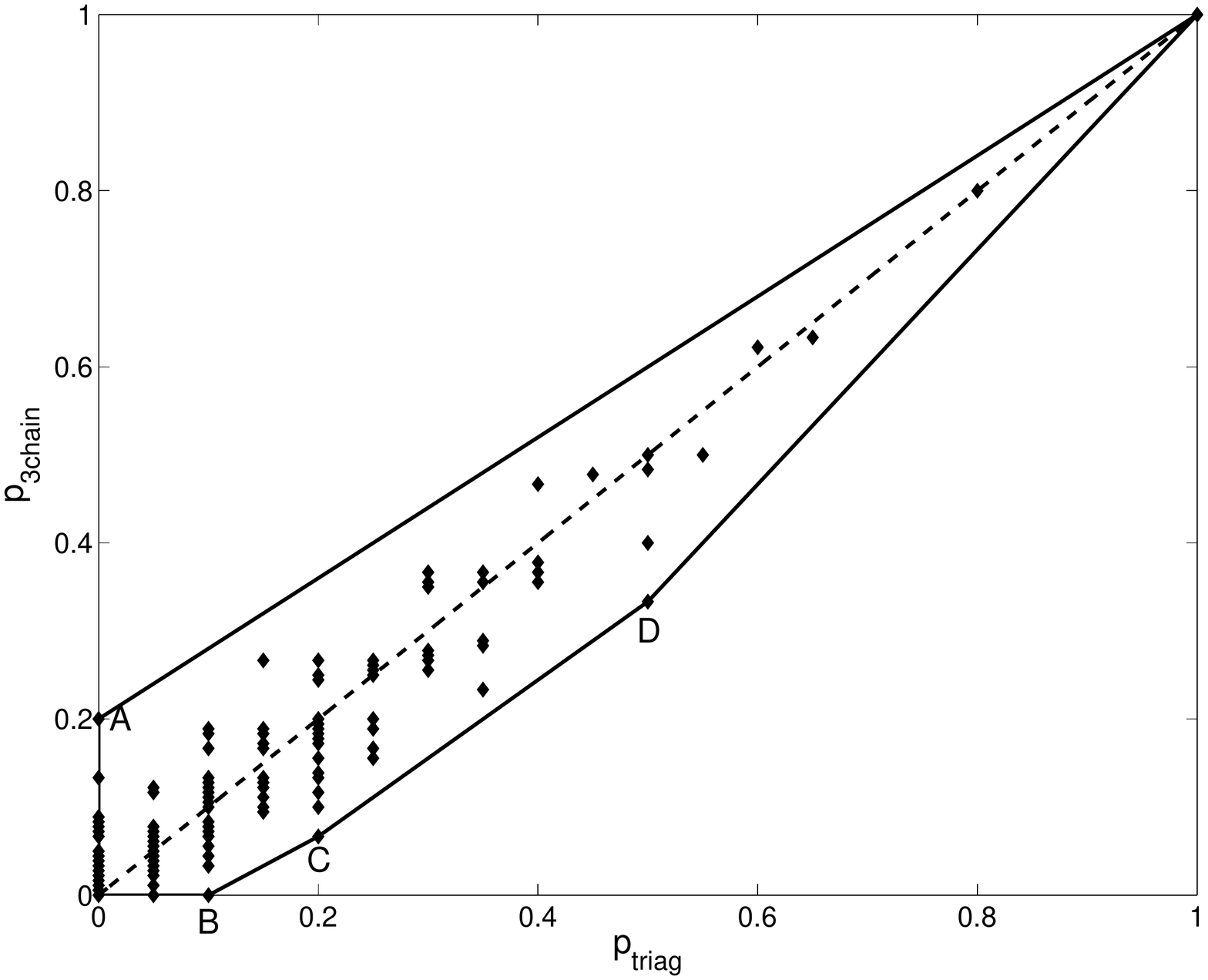}
\begin{minipage}[b]{0.25\textwidth}
\raisebox{10ex}{A} \includegraphics[width=0.42\textwidth]{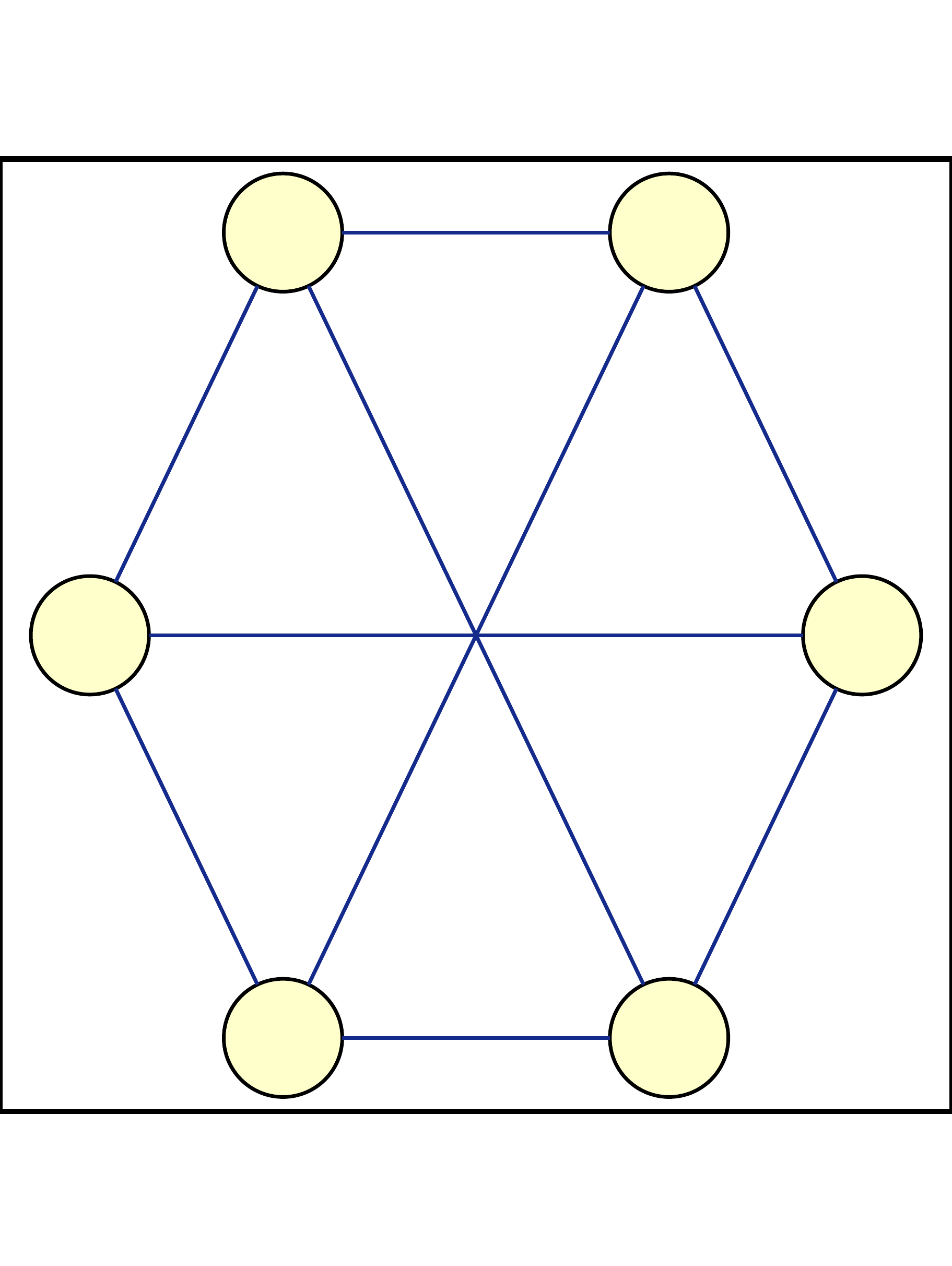} \\
\raisebox{10ex}{B} \includegraphics[width=0.42\textwidth]{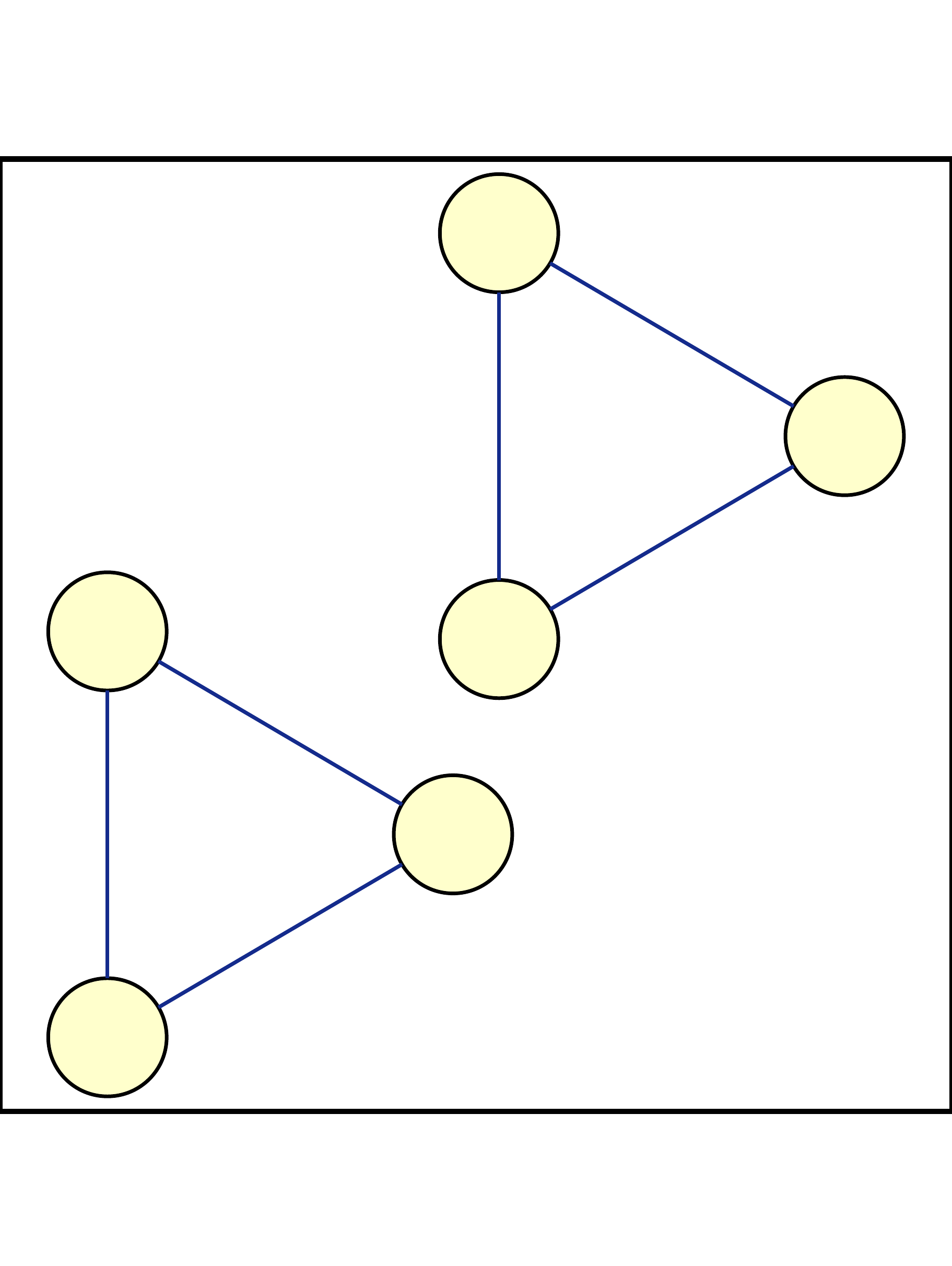} \\
\raisebox{10ex}{C} \includegraphics[width=0.42\textwidth]{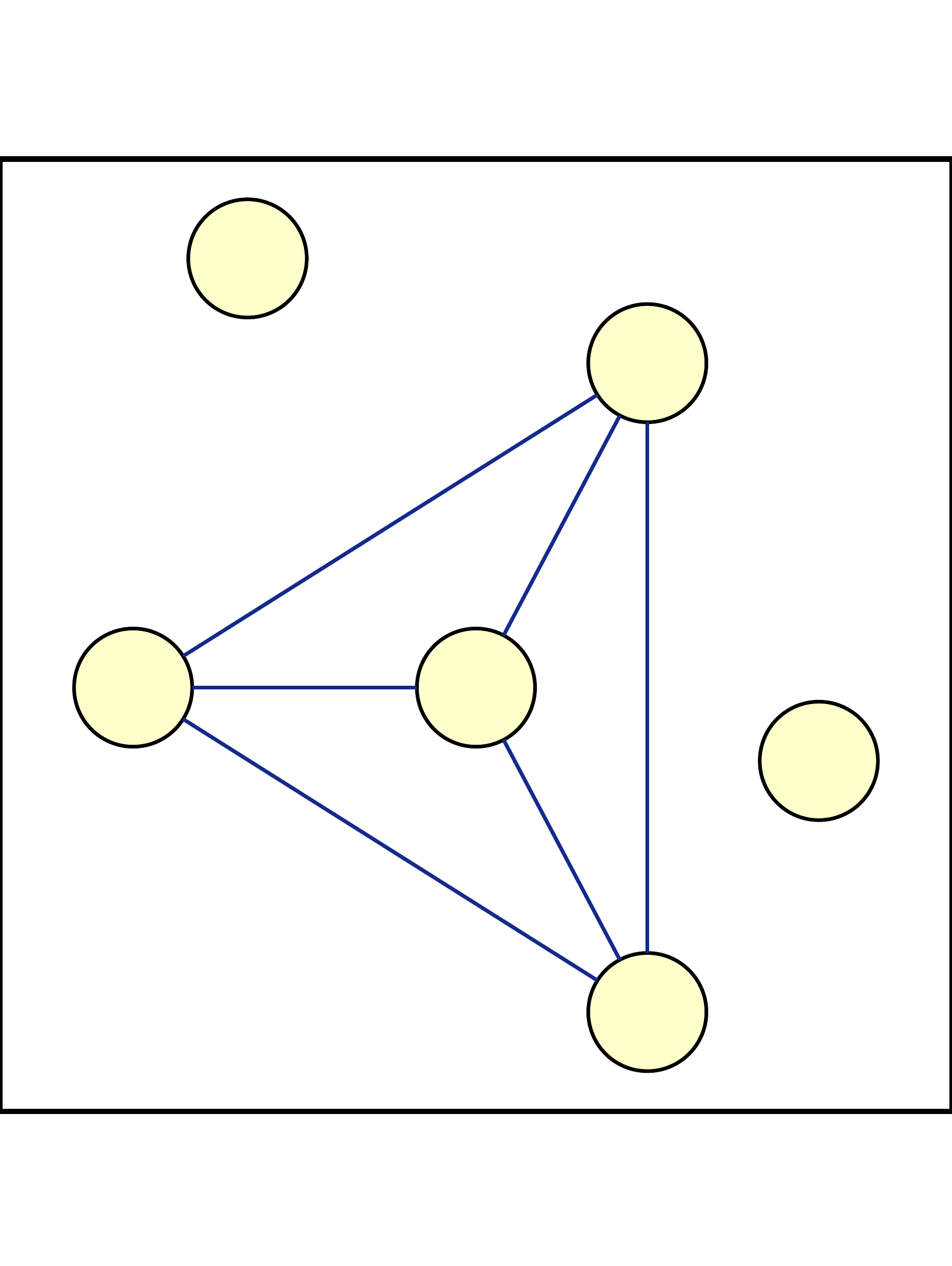}
\includegraphics[width=0.42\textwidth]{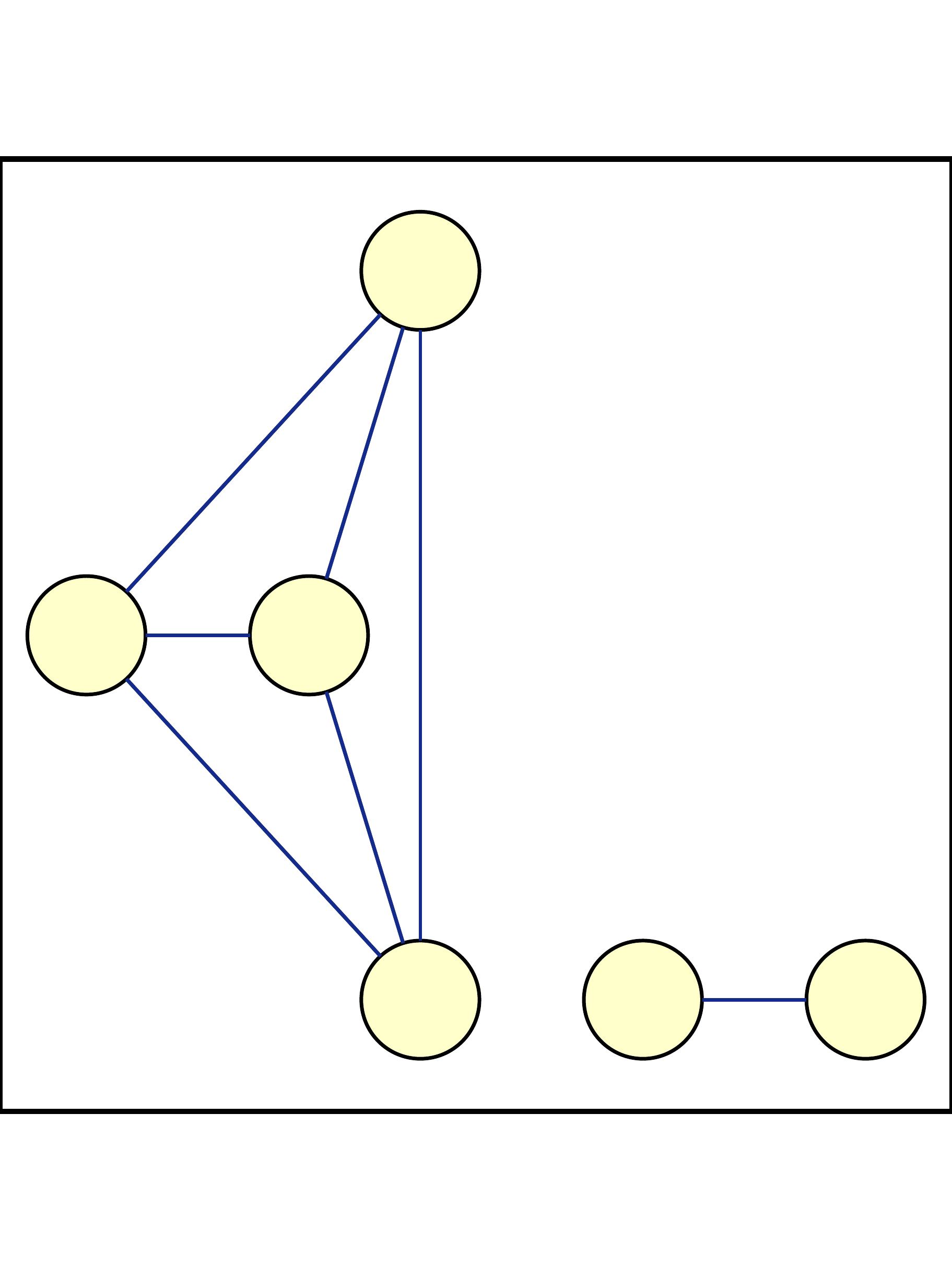} \\
\raisebox{10ex}{D} \includegraphics[width=0.42\textwidth]{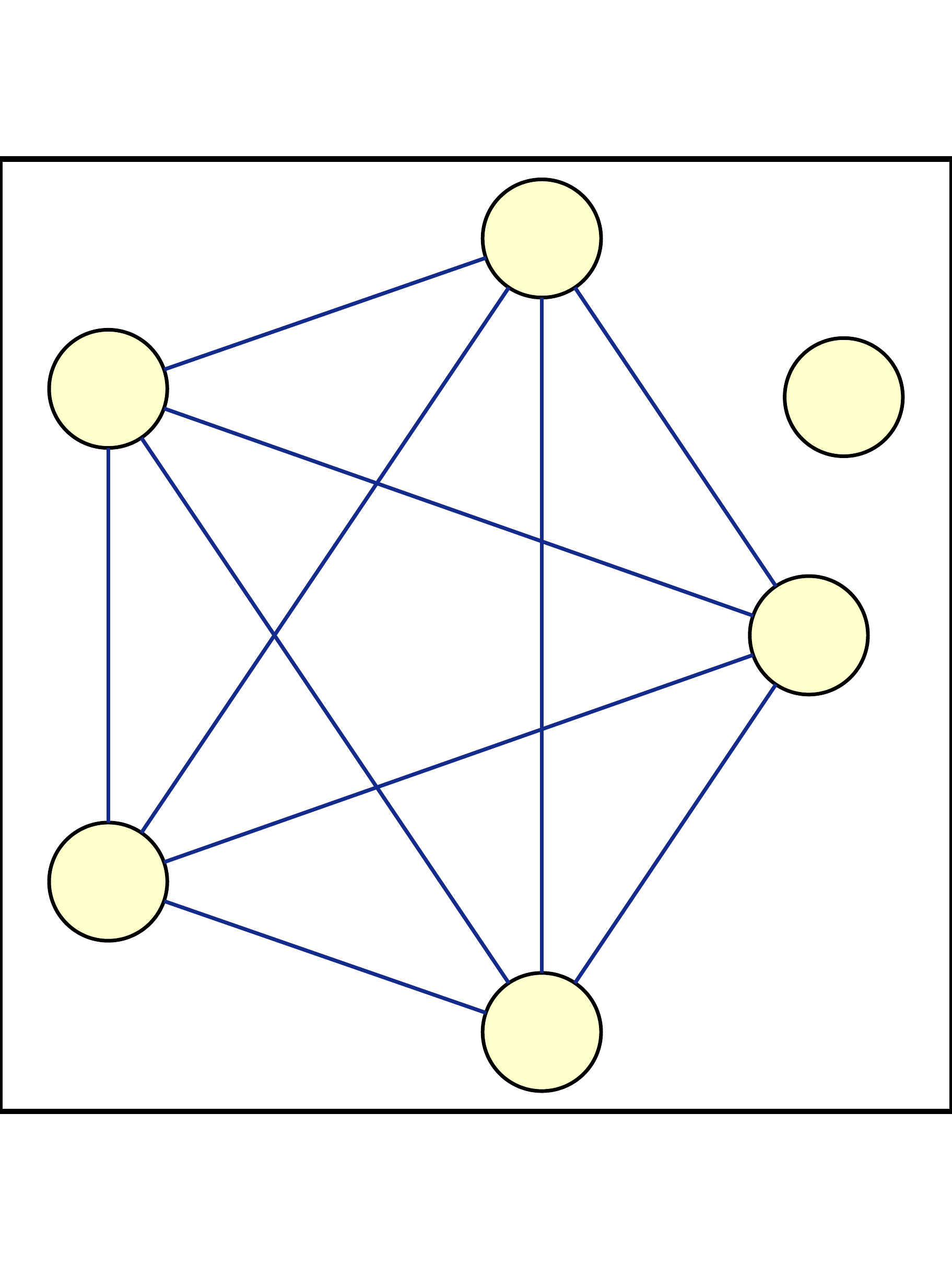}
\end{minipage}
\end{center}
\caption{Position of all graphs with $N=6$ nodes in the 
$(p_{\protect \three},p_{\protect \nine})$-plane. Graphs corresponding
to the extremal points of the convex hull are shown to the right.}
\label{fig:ninethree}
\end{figure}
\be
\Ha=c_{\nine} p_{\nine}+ c_{\three} p_{\three} 
\label{tt_model}
\ee Another simple exponential random graph model is given by
(\ref{tt_model}). By the same reasoning as above $c_{\three}$ and
$c_{\nine}$ should have opposite signs in order to have non-trivial
typical graphs. Fig.~\ref{fig:ninethree} shows again the region of admissible
expectation values $p_{\nine}$ and $p_{\three}$, all $6$-node graphs,
and the line $p_{\three}=p_{\nine}=p_{-}$ of the \ER
graphs. The minimal energy graphs can be easily understood in this
example. In the case of negative $c_{\nine}$ and positive $c_{\three}$
triangles are preferred. The minimal energy graphs are lying on the
lower boundary of the admissible region and are non-connected graphs
with fully connected components that could be considered as the ideal
case of a ``community structure'' (see for instance~\cite{Newman2004}). The size of the components depends on the concrete values of the parameters. If
the components are of different size, the graph is fully assortative, i.e. $r=1$. In the opposite case of positive $c_{\nine}$ and negative
$c_{\three}$ triangles are suppressed, and the minimal energy graph is a complete bipartite graph. The numbers of nodes in the two subsets are equal if the total number of nodes is even. Fig.~\ref{fig:ninethree}A shows this graph in the case of $6$ nodes. If the total number of nodes is odd, the numbers of nodes in the two subsets will differ by one. As a consequence the minimal energy graph in this case will be fully disassortative. This disassortativity is a consequence of the bipartiteness and the different size of the components, thus not very informative on its own. The same applies for
the observed assortativity in the first case that is also the
consequence of the very specific structure of these minimal energy
graphs. At the moment it is not clear to which extent it is possible
and reasonable to explain assortativity and disassortativity by a
catalog of paradigmatic structures that correspond to minimal energy
graphs in exponential random graphs model as in these simple examples.

\section{Discussion}
\label{sec:discussion}
We have presented a formalism that allows to study and quantify
systematically the structures of networks as statistical
dependencies. In particular, we showed how popular measures of network
structures such as the degree distribution, the cluster coefficient
and the assortativity coefficient could be expressed by subgraph
probabilities. This allows to situate graph ensembles with
predetermined values of these properties in the elements of the
hierarchy of exponential families \eqref{eq:amari-hierarchy} which
illuminates both their relationship and to which extent they specify
redundant information about the graph structure. Very often only a
single property is studied. For instance in~\cite{Newman2002}, a
random graph model with given degree distribution and additionally
given joint remaining degree distribution for connected links is
considered. This model allows for control of the degree of
assortativity, corresponding to a variation of $P(G)$ in one
direction. Depending on the exponential family $\mathcal{E}_k$ chosen, 
there are many other directions with non-vanishing
assortativity. Thus it remains unclear how relevant this particular
direction is. 

By identifying the subgraph counts as sufficient
statistics for exponential random graph models we also provide a link
to systematically incorporate motif analysis in the analysis of
network structures. A more detailed analysis of this aspect is beyond
the scope of this paper and will be presented elsewhere.
  

\begin{appendix}
\section{Expressing the assortativity coefficient by subgraph counts}
\label{sec:expr-assort-coeff}
Due to the homogeneity assumption all expectation values occurring in (\ref{equ:r2}) can be estimated as the average over all links in a given graph. Then the following relations to the subgraph counts are derived:
\begin{itemize}
\item $\langle d^{r,i} \rangle$:
\begin{eqnarray*}
\widehat{\langle d^{r,i} \rangle} & = & \frac{\sum_{i,j} a_{ij} d^{r,i}_{ij}}{\sum_{i,j} a_{ij}} \\
& = & \frac{\sum_{i,j} \left( a_{ij} \sum_{k \neq i,j} a_{ki} \right)}{\sum_{i,j} a_{ij}} \\
& = & \frac{\sum_{i,j \neq k} a_{ki} a_{ij}}{\sum_{i,j} a_{ij}} \\
& = & \frac{2 \cdot n_{\two}}{2 \cdot n_{-}} = \frac{n_{\two}}{n_{-}}
\end{eqnarray*}
Since the average is performed with respect to all links, nodes with high degree get high weights.
Thus, even though $d^{r,i}_{ij} = d_i - 1$, $\frac{1}{\sum_{i,j} a_{ij}} \sum_{i,j} d^{r,i}$ 
is not equal to $\frac{1}{N} \sum_i d_i - 1$!
\item $\langle d^{r,i} d^{r,j} \rangle$:
\begin{eqnarray*}
\widehat{\langle d^{r,i} d^{r,j} \rangle} & = & \frac{\sum_{i,j} a_{ij} d^{r,i}_{ij} d^{r,j}_{ij}}{\sum_{i,j} a_ij} \\
& = & \frac{\sum_{i,j} \left( a_{ij} \sum_{k \neq i,j} a_{ki} \sum_{l \neq i,j} a_{jl} \right)}{\sum_{i,j} a_ij} \\
& = & \frac{\sum_{i,j \neq k} a_{ki} a_{ij} a_{jk}  + \sum_{i,j \neq k \neq l} a_{ki} a_{ij} a_{jl}}{\sum_{i,j} a_ij} \\
& = & \frac{3 \cdot n_{\nine} + n_{\three}}{n_{-}}
\end{eqnarray*}
\item $\langle (d^{r,i})^2 \rangle$:
\begin{eqnarray*}
\widehat{\langle (d^{r,i})^2 \rangle} & = & \frac{\sum_{i,j} a_{ij} \left( \sum_{k \neq i,j} a_{ki} \right)^2}
{\sum_{i,j} a_{ij}} \\
& = & \frac{\sum_{i,j} \left( a_{ij} \sum_{k \neq i,j} a_{ki} \sum_{k' \neq i,j} a_{k'i} \right)}
{\sum_{i,j} a_{ij}} \\
& = & \frac{\sum_{i,j \neq k} a_{ij} \overbrace{a_{ki}^2}^{= a_{ki}} + \sum_{i,j \neq k \neq k'} a_{ij} a_{ki} a_{k'i}}
{\sum_{i,j} a_{ij}} \\
& = & \frac{2 \cdot n_{\two} + 6 \cdot n_{\four}}{2 \cdot n_{-}} \\
&=& \frac{n_{\two} + 3 n_{\four}}{n_{-}}
\end{eqnarray*}
\end{itemize}
Hence the assortativity expressed in subgraph counts is
\besn
\widehat{r^2}&=&\frac{3 \cdot n_{\nine} + n_{\three}-\frac{n_{\two}^2}{n_{-}}}{\left( n_{\two} + 3 \cdot n_{\four}- \frac{n_{\two}^2}{n_{-}} \right)} \\
&=& \frac{n_{-} \left( 3 \cdot n_{\nine} + n_{\three} \right)-n_{\two}^2}{\left( n_{-} n_{\two} +3 n_{-} n_{\four} - n_{\two}^2 \right)} \\
&=&\frac{\frac{n_{-}}{n_{\two}} \left(\frac{3 n_{\nine}}{n_{\two}}+\frac{n_{\three}}{n_{\two}} \right)-1}{\frac{n_{-}}{n_{\two}} \left(\frac{3 n_{\four}}{n_{\two}}+1 \right)-1} \;.
\eesn 

\end{appendix}

\bibliographystyle{spmpsci}      
\bibliography{networks}   

%
%

\end{document}